\documentclass[twocolumn,pra,showpacs,amsmath,amssymb]{revtex4-1}
\usepackage{graphicx}
\usepackage{bm}
\usepackage{epstopdf}
\usepackage{epsfig}
\usepackage{dsfont}
\begin{document}

\title{Analytic quantification of the singlet nonlocality for the first Bell's inequality}
\author{Fernando Parisio}
\email[]{parisio@df.ufpe.br}
\affiliation{Departamento de
F\'{\i}sica, Universidade Federal de Pernambuco, Recife, Pernambuco
50670-901 Brazil}

%\date{\today}

\begin{abstract}
Recently an alternative way to quantify Bell nonlocality has been proposed 
[Phys. Rev. A {\bf 92}, 030101(R) (2015)]. In this work we further develop this concept, the volume of violation, and analytically calculate its value for the spin-singlet state with respect to the settings of the first Bell's inequality. These settings correspond to
three directions in space, or three arbitrary points on the unit sphere. It is shown that the triples of directions that lead 
to violations in local causality correspond to $1/3$ of all possible configurations. From the perspective of quantum communications, this means that two distant parties that were capable of align their measurements in one direction only (the remaining direction in each site being random), have a probability of about 33.3$\%$ to be able to certify their entanglement.
\end{abstract}
\maketitle

\section{Introduction}
\label{intro}
The fact that Bell inequalities are violated in increasingly convincing experiments \cite{hensen,giustina,shalm} led the majority of the physics community to accept that nature is able to display nonlocal behaviors. In addition, the violations are such that the predictions of quantum mechanics are strictly observed, and, the so far unbeaten theory is itself said to be nonlocal. But how nonlocal? 

Although one feels compelled to quantify any resource that becomes relevant in both, basic science and technology, this program is very hard when one deals with nonlocality. Let us begin with a remark on entanglement, a related but distinct concept. Supposing that the entanglement content of multipartite mixed systems can be characterized by a finite set of numbers, it is not too risky to say that most of us, if not all, would agree on one point: These numbers should be a property of the quantum state alone. This means that in a hypothetical full-fledged theory of entanglement quantification, no context descriptions should be necessary whatsoever. For this reason entanglement measures would not be influenced by the fact that quantum mechanics is contextual. 

Not even this starting point is clear in the case of nonlocality. Many consider that also in this case an ultimate measure should depend only on the state. But this a less defendable position. Nonlocality by its very definition refers to space-time, whose features do not enter in Bell's inequalities for intrinsic degrees of freedom (spin, polarization, etc). Most importantly, even from a theoretical perspective, to asses nonlocality one must consider some kind of measurement, which makes unavoidable to face the contextual character of quantum mechanics. Thus, it is not unreasonable to suppose that the nonlocality of a quantum state should display some dependence on the context (e. g., on the number and orientations of Stern-Gerlach apparatuses being employed). One possibility in this direction is to quantify the amount of nonlocality with respect to the setup required to investigate violations of a particular Bell's inequality $I \leq \xi_c$, where $\xi_c$ is the limit imposed by local causality. A common procedure to induce an ordering in the Hilbert space is as follows. Consider a particular state $\varrho$ and the Bell function associated to the chosen inequality,
\begin{equation}
\nonumber
I(\varrho;x_1,x_2,\dots, x_n)\;,
\end{equation}
where $\{x_i\}$ is the set of parameters that can be varied in the measurement (it is important to stress that we are restricting our analysis to rank-1 projective measurements). Then, a state $\rho$ is said to be more nonlocal than $\sigma$ if $I_{\rm max}(\rho)>I_{\rm max}(\sigma)>\xi_c$. The maximum of the Bell function is obtained by visiting all setting parameters. 
This relation certainly means that $\rho$ is more correlated than $\sigma$, both being correlated beyond any classical system could be.

But, does it make sense to use this criteria as a measure of nonlocality? 
Consider that there are quantum states for which $\xi_c<I\le\xi_q$, where $\xi_q$ is the maximum value of $I$ according to quantum mechanics. For definiteness, suppose that $\xi_q=\sqrt{2}\xi_c$ and consider two states such that
\begin{equation}
\nonumber
I_{\rm max}(\rho)=\sqrt{2}\xi_c\;, \;\;I_{\rm max}(\sigma)=\xi_c+\epsilon\;,
\end{equation}
with $\epsilon$ being a positive number that can be made arbitrarily small.
Assume, for simplicity, that these values are attained for the same set of parameters (e. g., a particular set of directions). For this fixed situation what makes one states that $\rho$ is more nonlocal? Perhaps, since the violation by $\sigma$ is weaker, the subsystems in this case should be farther away in comparison to those described by $\rho$ in order to enable violations. Another possibility is to think of superluminal signaling. Perhaps, the minimal signal velocity for $\rho$ must be larger than the analogous quantity for $\sigma$: $u_{\rho}>u_{\sigma}>c$, where $c$ is the speed of light, and $u$ is the minimum signal velocity for each state. However, none of these points can be sustained by facts. In particular, it is not true that $u_{\sigma}=c+O(\epsilon)$, no matter how small is $\epsilon$. Since the distance between the subsystems do not enter in the inequalities and can be arbitrarily large, one would need $u_{\sigma}\rightarrow \infty$ even if $\epsilon \rightarrow 0$. The truth is that, for the fixed set of parameters, the two states are equally nonlocal, because the action at a distance required to explain the quantum correlations must be equally ``spooky'' in the two cases. 

It is instructive to look at this question from the opposite perspective. Consider two non-violating states $\rho'$ and $\sigma'$. Is it reasonable to say that, because $I_{\rm max}(\rho')<I_{\rm max}(\sigma')\le\xi_c$ regarding some particular setting, $\rho'$ is more {\it local} than $\sigma'$? Of course not. If two states are local, nothing else needs to be added. There are no gradations of locality.

Note carefully that it is not being implied that $I_{\rm max}(\rho)>I_{\rm max}(\sigma)>\xi_c$ has no meaning with regards to nonlocality. To see this, let us vary by a small amount some of the parameters so that the violations are no longer maximal: 
\begin{equation}
\nonumber
I_{\rm submax}(\rho)=\sqrt{2}\xi_c-2\epsilon\;, \;\;I_{\rm submax}(\sigma)=\xi_c-\epsilon\;.
\end{equation}
Now, state $\sigma$ is local while $\rho$ is still nonlocal, for sufficiently small $\epsilon$, with respect to the new setting parameters. So, it may be appropriate to look at violations for all possible parameters.
What is proposed in reference \cite{fonseca} is that, for a fixed set of parameters, there should be no gradations of nonlocality, and that it should be assessed only when all experimental situations are considered. In this context, the state $\rho$ is more nonlocal than $\sigma$ if the former violates local causality, no matter by what extent, for a larger number of experimental configurations than the latter. This consideration led to the definition of the volume of violation
\begin{equation}
V(\varrho)=V(\varGamma_{\varrho})= \int_{\varGamma_{\varrho}}d^n x \;,
\end{equation}
where $\varGamma$ is the subset of ${\cal X }=\{ x_i \}$ containing al violating configurations and $d^n x=\mu(x_1,\dots, x_n)dx_1\dots dx_n$.
The measure $\mu$ is such that every configuration is equally important as will become clear in the example of the next section. Note that, according to the above definition, $V(\rho')=V(\sigma')=0$, since $\varGamma_{\rho'}=\varGamma_{\sigma'}=\emptyset$, meaning, as expected, that all non-violating states are equally local. We stress that this definition aims at quantifying the nonlocality of a physical situation, i. e., of a state $\rho$ under the specific measurements required to test some fixed Bell's inequality.

The nonlocal content of a state under a certain kind of measurement, as described by the volume of violation, has been applied to a problem that became 
known as the ``anomaly'' in the nonlocality \cite{acin,methot}. It consists in the fact that the Collins-Gisin-Linden-Massar-Popescu (CGLMP) inequality \cite{collins} is maximally violated by a state that is not maximally entangled for two entangled qutrits. It turns out that the supposed anomaly disappears when the volume of violation in used, that is, $V$ attains its maximum for the maximally entangled state. Of course, nonlocality and entanglement are distinct quantities, e. g., the latter is more fragile against noise than the former. What the referred result seems to indicate is that the difference is not as evident as some results, like the anomaly, initially suggested.

The integrations involved in the determination of the volume of violation for two three-level systems under CGLMP measurements required a fully numeric approach, thus, not being very illustrative on how to determine these volumes in detail. In what follows we show that, at least in the simplest case, that of a spherically symmetric state subjected to a test of the first Bell's inequality, the volume of violation can be computed analytically. 

\section{The simplest case}

In his milestone paper, Bell showed that the predictions of quantum mechanics would be, in certain testable situations,
in conflict with any theory preserving local causality \cite{bell}. This fundamental conflict appeared in the form of 
an inequality for correlations between spin-1/2 measurements made by two parties $A$ and $B$ sufficiently far apart. 
The relation that should be satisfied by any theory compatible with local causality reads
\begin{equation}
\label{bell1}
|E({\bf a},{\bf b})-E({\bf a},{\bf c})|\le 1+E({\bf b},{\bf c})\;,
\end{equation}
where $E({\bf a},{\bf b})$ is the average value of the product of the result of spin measurements obtained by $A$ along direction ${\bf a}$ and by $B$ along direction ${\bf b}$. Denoting the possible outcomes by $+1$ and $-1$ we get $E({\bf a},{\bf b})=P(+1,+1)+P(-1,-1)-P(+1,-1)-P(-1,+1)$, where $P(+1,+1)$ is the probability for results $+1$ in $A$ and $+1$ in $B$, and so on.
%
%\section{Singlet volume of violation for the first Bell's Inequality}

\begin{figure}
\label{fig1}
\includegraphics[width=7cm]{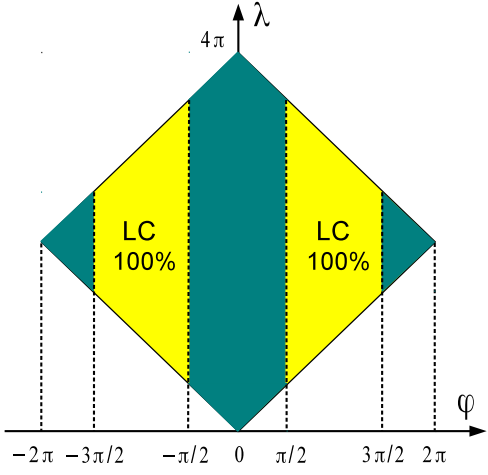}
\caption{In the lighter portions of the $\varphi$-$\lambda$ domain the inequality is trivially satisfied. In the darker regions local causality (LC) may be violated.}
\end{figure}

\begin{figure}
\label{fig2}
\includegraphics[width=9cm]{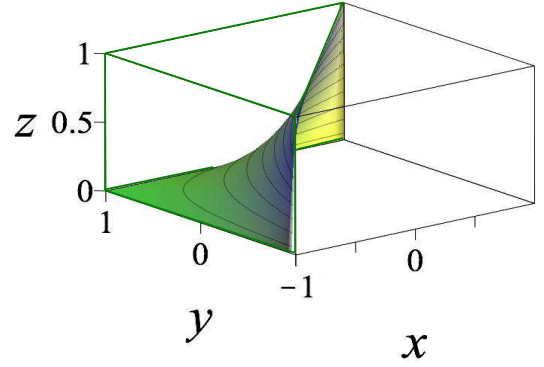}
\caption{The concave portion in the left part of the cell corresponds to the violation region. The variables are $x=\cos \theta_b$, $y=\cos \theta_c$, and $z=\cos^2 (\varphi)$. All variables are dimensionless.}
\end{figure}

Hereafter we will address the singlet state
\begin{equation}
\label{sing}
|\Psi_s\rangle=\frac{1}{\sqrt{2}}(|+1,-1\rangle-|-1,+1\rangle)\;.
\end{equation}
Perhaps the most important point about this vector is that it is unnecessary to specify which basis is being employed to represent it. The singlet has the invariant form (\ref{sing}) with $|\pm1\rangle$ being the eigenstates of $\boldsymbol{\sigma}\cdot {\bf n}$ for {\it any} direction ${\bf n}$, with $\boldsymbol{\sigma}=(\sigma_x,\sigma_y,\sigma_z)$, $\sigma_i$ denoting the Pauli matrices. This greatly facilitates our task. Since there are no privileged directions, without loss of information, we can assume that one of the measurement directions is fixed in the space, say, ${\bf a}=(0,0,1)$. Furthermore, the correlations assume a particularly simple form for the singlet, namely, $E({\bf a},{\bf b})=-{\bf a} \cdot {\bf b}$. The inequality (\ref{bell1}) becomes
\begin{equation}
\label{bell1b}
|\cos \theta_c-\cos \theta_b| \le 1-\sin \theta_c \sin \theta_b \cos \varphi-\cos \theta_c \cos \theta_b
\end{equation}
where $\theta_c$, $\theta_b$ $\in [0,\pi]$ are the azimuthal angles of ${\bf c}$ and ${\bf b}$, respectively, and $\varphi=\phi_c-\phi_b$ $\in [-2\pi,2\pi]$ is the difference between the corresponding polar coordinates. The complementary variable $\lambda=\phi_c+\phi_b \in [0,4\pi]$ does not enter in the inequality. We, thus, have a four-dimensional space of parameters, which we denote by ${\cal X}$. We intend to calculate volumes in this space, in particular the volume of the subset $\varGamma_s$ (for the singlet) which corresponds to setting parameters that lead to violations in (\ref{bell1b}). The measure of integration must be neutral in the sense that any configuration is equally relevant, that is
\begin{eqnarray}
\nonumber
d^4x=d\Omega_c d\Omega_b=\sin \theta_c \sin \theta_b d\theta_c d\theta_b d\phi_c d\phi_b \\
=\frac{1}{2} \sin \theta_c \sin \theta_b d\theta_c d\theta_b d\varphi d\lambda \;,
\end{eqnarray}
where, in the first line the polar integrations are over a square with side length of $2\pi$ and in the second line the integrations are also over a square domain, see Fig \ref{fig1}, delimited by $\lambda=4\pi-|\varphi|$, and $\lambda=|\varphi|$. This leads to a total volume of
\begin{equation}
V({\cal X})=16\pi^2\;,
\end{equation}
corresponding to the product of two solid angles, as it should be (we are skipping the trivial solid angle corresponding to rotations of ${\bf a}$). Our objective is to calculate the volume of the set $\varGamma_s$, or more importantly, the ratio between the volumes of $\varGamma_s$ and ${\cal X}$, that we define as the relative volume of violation $v(\varGamma_s)$. For half of the possible configurations we have $|\cos \theta_c-\cos \theta_b|=\cos \theta_c-\cos \theta_b$. This situation ($\theta_c \le \theta_b$) obviously corresponds to one half of the violating configurations and, therefore, one can focus on it, doubling the result in the end. In addition, since $\sin \theta_c$ and $\sin \theta_b$ are non negative in $[0, \pi]$ we can write $\sin \theta_c \sin \theta_b=+\sqrt{(1-\cos^2 \theta_c)(1-\cos^2 \theta_b)}$. For $\theta_c\ne0$ and $\theta_b\ne\pi$, the inequality assumes the more symmetrical form
\begin{equation}
\nonumber
%\label{bell1c}
\sqrt{(1+x)(1-y)}\ge \cos \varphi \sqrt{(1-x)(1+y)}\;,
\end{equation}
where we defined the one-to-one relations $x=\cos \theta_b$, $y=\cos \theta_c \in [-1, 1]$. If we split $\varphi$ in the intervals shown in Fig. \ref{fig1}, covering $[-2\pi,2\pi]$, then, in the lighter regions the inequality is trivially fulfilled $(\cos \varphi\le 0)$. In the remaining regions, where $\cos \varphi \ge 0$, it is harmless to write
\begin{equation}
\label{bell1d}
(1+x)(1-y)\ge \cos^2 \varphi \;(1-x)(1+y)\;,
\end{equation}
Accordingly, the surfaces delimiting the violating region are given by $\cos^2 \varphi =(1+x)(1-y)/(1-x)(1+y)$ and the planes $x=-1$, $y=1$, and $\cos^2 \varphi=1$. It is convenient to introduce the variable $z(\varphi)=\cos^2(\varphi) \in [0,1]$ which defines a one-to-one relation in {\it each} of the four dark parts in Fig. \ref{fig1}, namely, 
\begin{equation}
\label{intervals}
[-2\pi,-3\pi/2],\; [-\pi/2,0],\; [0,\pi/2],\; [3\pi/2,2\pi]\;.
\end{equation}
In Fig. \ref{fig2} we show one of these four tridimensional cells.
The volume of violation of the singlet state is, then, given by 
\begin{eqnarray}
\nonumber
V(\varGamma_s)=2\times\int_{\varGamma_{y>x}} d\mu\\
= \int_{-2\pi}^{2\pi} \int_{|\varphi|}^{4\pi-|\varphi|} \int_{-1}^{+1}\int_{y_{\varphi}(x)} ^{+1}dy dx d\lambda d\varphi\;,
\label{int}
\end{eqnarray}
where $\varGamma_{y>x}$ is the restriction of $\varGamma_{s}$ to $y>x$, and
\begin{equation}
\nonumber
y_{\varphi}(x)=\frac{(1+x)-z(\varphi)\;(1-x)}{(1+x)+z(\varphi)(1-x)}\;,
\end{equation}
which gives a family of curves in the $x-y$ plane as $\varphi$ varies. Note that for $z=1$ we have $y=x$ as can be seen on the top of Fig. \ref{fig2}, while for $z=0$ we get $y=1$. The volume in Eq. (\ref{int}) can be written as
\begin{equation}
\nonumber
V(\varGamma_{s})
=\int_{-2\pi}^{2\pi} (4\pi-2|\varphi|) A(\varphi) d\varphi\;,
\end{equation}
where $A(\varphi)$ is the area in the $x-y$ plane delimited by $y_{\varphi}(x)$ and $y=1$. We only need to consider the variable $\varphi$ in the regions (\ref{intervals}). The symmetric intervals with respect to $\varphi=0$ are equivalent and, thus, we get
\begin{equation}
\nonumber
V(\varGamma_{s})=2\left( \int_{0}^{\pi/2}+\int_{3\pi/2}^{2\pi}\right)(4\pi-2\varphi )A(\varphi)d \varphi\;.
\end{equation}
We now conclude the change of variables $z=\cos^2(\varphi)$. In the first interval ($\varphi \in [0,\pi/2] \rightarrow z \in [1,0]$) $\varphi=\arccos(\sqrt{z})$ while in the second interval ($\varphi \in [3\pi/2,2\pi] \rightarrow z \in [0,1]$) $\varphi=2\pi-\arccos(\sqrt{z})$.
We, therefore, have
\begin{eqnarray}
\nonumber
d\varphi=\nu \frac{dz}{2\sqrt{z(1-z)}} \;,
\end{eqnarray}
with $\nu=+1$ in the first interval and $\nu=-1$ in the second interval. Gathering all this together we obtain
\begin{equation}
\label{int3}
V(\varGamma_{s})=4\pi \int_{0}^{1}\frac{A(z)dz}{\sqrt{z(1-z)}}\;.
\end{equation}
It is easy to find
\begin{equation}
\nonumber
A(z)=2-\frac{2}{(1-z)^2}\left[2z\ln(z)+(1-z^2)\right]\;,
\end{equation}
which turns out to be a quite inconvenient representation of $A$, since its direct insertion in integration (\ref{int3}) leads to divergencies. The second and third terms in the previous equation lead to an indetermination of type ``$\infty-\infty$'' . This can be dealt with by expanding the logarithm around $z=1$. We get
\begin{equation}
A(z)=4z\sum_{n=0}^{\infty}\frac{(1-z)^n}{n-2}\;,
\end{equation}
which is explicitly finite under (\ref{int3}). In addition, since it is a convergent series in the range of definition of $z$, one can interchange the ordering of summation and integration, leading to
\begin{eqnarray}
\label{int4}
\nonumber
V(\varGamma_{s})=16\pi \sum_{n=0}^{\infty}\frac{1}{n+2}\int_{0}^{1}z^{1/2}(1-z)^{n-1/2} dz\\
\nonumber
=8\pi^{3/2}\sum_{n=0}^{\infty}\frac{\Gamma(n+1/2)}{\Gamma(n+3)}\;,
\end{eqnarray}
where $\Gamma$ denotes the gamma function. The infinite series can be written in terms of a hypergeometric function as $[\Gamma(1/2)/2]\times\,_2F_1(1/2,1;3;1)$, which gives $2\sqrt{\pi}/3$ and, thus
\begin{equation}
\label{end}
V(\varGamma_s)=\frac{16\pi^2}{3}\Rightarrow v(\varGamma_s)\equiv\frac{V(\varGamma_s)}{V({\cal X})}=\frac{1}{3}\;.
\end{equation}
Therefore, the relative volume of violation of the spin-1/2 singlet state is 1/3 with respect to the first Bell's inequality. 
\section{concluding remarks}
By reasoning that the numeric value of a Bell function should be taken as a witness, rather than as a quantifier of nonlocality (and of locality), a new measure has been defined \cite{fonseca}. In this paper we showed that this quantity can be analytically calculated for the singlet state in the setup necessary to investigate the first Bell's inequality. Our result means that
if we randomly pick three directions in space, without any bias, then, the probability that the selected configuration will lead to a violation in local causality is $1/3$. This may serve as an initial test to numerical procedures aiming at the calculation of volumes of violation associated to more complex states and contexts.

From a quantum communications perspective, our result means that two distant parties that were able to align their measurement apparatuses in {\it one} direction only, the remaining direction in each site being random, have a probability to certify their entanglement of about 33.3 $\%$ (for the singlet). 
This can be understood as follows. For the first Bell's inequality one of the measurement directions, in our notation ${\bf b}$ [see Eq. (\ref{bell1})], is present in the arguments referring to $A$ and $B$, thus, they have to agree on that direction in advance. Had we considered the Clauser-Horne-Shimony-Holt (CHSH) inequality \cite{chsh}, then no alignment would be required whatsoever and the probability to certify entanglement would be much lower. According to our numerical results, it is about $5\%$. Conversely, it is well known that the first Bell's inequality derives from CHSH, which involves four independent directions (say ${\bf a},{\bf a'}, {\bf b}, {\bf b'}$), when we set ${\bf a'}={\bf b'}$.
These same issues were addressed in \cite{liang,wallman}, where the authors employ the very concept of volume of violation (but not in the context of measures of nonlocality). The authors of reference \cite{fonseca} became aware of these publications very recently.

An interesting perspective is to extend the concept of volume of violation to positive-operator-valued measurements (POVM's), especially for inequalities where these more general measurements are required in order to attain maximal violation \cite{vertesi}. In this context a physically relevant question arises: Given a fixed inequality and a set of POVM's, does the volume of violation depend on the particular Naimark realization that is employed?

Even with the simplifications brought by spherical symmetry, the volume of violation regarding the settings of the CHSH inequality amounts to a five-dimensional integration, which, so far, we were not able to carry out analytically. The main goal is to compare this volume to that of the Popescu-Rohrlich box \cite{pr}. Since we are arguing that the numeric value of a Bell function is being overrated, the fact that these probability boxes can yield $I=4>2\sqrt{2}$ for the CHSH inequality, does not necessarily mean that they are more nonlocal than quantum mechanics, at least in the framework of our proposal.

\begin{acknowledgments}
The author thanks A. P. Pereira da Costa and E. A. Fonseca for useful discussions on this problem. The author would like to thank the anonymous referee for bringing to his knowledge the references \cite{liang,wallman}.
This work received financial support from the Brazilian agencies Conselho Nacional de Desenvolvimento Cient\'{\i}fico e Tecnol\'ogico (CNPq), Coordena\c{c}\~ao de Aperfei\c{c}oamento de Pessoal de N\'{\i}vel Superior (CAPES), and Funda\c{c}\~ao de Amparo \`a Ci\^encia e Tecnologia do Estado de Pernambuco (FACEPE).
\end{acknowledgments}

\end{document}